\documentclass[12pt]{article}
\usepackage{amssymb}
\usepackage{graphicx}
\usepackage{epstopdf}
\usepackage{sw20aip}

\input{tcilatex}
\begin{document}

\date{February 5, 2008}
\title{Rise and premature fall of the old quantum theory}
\author{Manfred Bucher \\
Physics Dept., Cal. State Univ., Fresno, CA 93740-8031, USA}
\maketitle

\begin{abstract}
The old quantum theory of Bohr and Sommerfeld was abandonned for the wrong
reason. Its contradictions were caused not by the orbit concept but by a
mental barrier---the inconceivability that an electron might collide with
the atomic nucleus. Removing that barrier resolves the theory's main
failures---incorrect orbital momenta, $He$ atom, $H_{2}^{+}$ molecule ion.
The inclusion of electron oscillations through the nucleus---a concept
called ``Coulomb oscillator''---renders the old quantum theory consistent
with quantum mechanics (although devoid of wave character). The triple
success of the Bohr-Sommerfeld model is its correct description of the $H$
atom (and one-electron ions) concerning (1) the energy levels $E_{nl}$, (2)
the orbital angular momenta $L_{nl}$---if corrected as $L_{nl}^{2}=l(l+1)%
\hbar ^{2}$and with the Coulomb oscillator included---and (3) the orbits'
space quantization---with $(L_{nl})_{z}=$ $m_{l}\hbar $. These achievements
are succinctly represented by the principal, angular and magnetic quantum
numbers ($n,$ $l,$ $m_{l}$) and visualized by orbital ellipse
geometry---major axis, vertex curvature, and tilt angle, respectively. Orbit
geometry also accounts for the average orbit size. Moreover, the Coulomb
oscillator provides a natural explanation of (isotropic) hyperfine
interaction. The shortcomings of the old quantum theory lie in its neglect
of three properties of particles---their spin, their wave nature and their
quantum statistics. These deficiencies notwithstanding, the visual appeal of
the Bohr-Sommerfeld model remains a pedagogical asset to complement the
abstract character of quantum mechanics.
\end{abstract}

\section{INTRODUCTION}

Quantum theory made its debut in the fall of 1900 when Max Planck, professor
at the University of Berlin, succeeded in deriving a formula for the
spectral distribution of thermal radiation that agreed precisely with
experiment (Planck 1900). The formula was based on the unconventional, and
(at the time) unjustified assumption that electromagnetic radiation at any
(visible or invisible) spectral color can be absorbed or emitted only in
tiny portions of energy, called ``quanta.'' Planck's formula---later called
``Planck's law''---resolved the long-standing puzzle of why all (``black'')
bodies at the same temperature have the same thermal color. Fostered by
advances in spectroscopy and thermodynamics, Gustav Kirchhoff, a previous
professor at Berlin, had discovered, a half-century earlier, this universal
relation (Kirchhoff 1859). Kirchhoff's discovery was recognized as being of
fundamental rather than technical importance---something akin to Newton's
universal law of gravity. Accordingly, much experimental and theoretical
effort was devoted to thermal-radiation spectroscopy. With Planck's
successful resolution of a problem of universal scope, quantum theory, while
still at the margins of physics, acquired a touch of foundational
significance. More importantly, the theory's underlying concept of stepwise
change introduced a new paradigm to the description of nature. The previous
paradigm---continuous change---had gained universal credibility because of
its basic role in the greatest triumph of mathematics: the calculus.

The second step in the advance of quantum theory occurred in 1905 when
iconoclast Albert Einstein, still patent clerk in Bern, used Planck's
quantum concept to explain the obscure photoelectric effect (Einstein 1905).
Einstein assumed that light consisted of ``light quanta''---later called
``photons''---that would hit electrons like bullets. Both Planck and
Einstein assumed a universal ratio between photon energy and radiation
frequency, $\epsilon /f=h$, later called Planck's constant. They differed,
though, in its interpretation. Planck regarded merely the radiative emission
and absorption process as quantized, but not, like Einstein, radiation 
\textit{per se}. In contrast to Planck's resolution of a universal problem
(thermal radiation), Einstein's contribution appeared less significant---an
attempt to explain an experimental oddity. Einstein invoked Planck's quantum
theory in two other cases. One was his theory of the specific heat of solids
at low temperatures (Einstein 1907). In the other case, Einstein calculated
energy fluctuations of radiation and found that they consisted of both wave
and particle contributions (Einstein 1909).

With its next appearance, however, quantum theory moved from the margins of
physics to center stage when it addressed the most important question of the
day: the structure of the atom. The notion of atoms had already appeared in
some natural philosophy of ancient Greece---a speculation vindicated only
two and a half millennia later. However, from the late 1700s on, the concept
of atoms had attracted serious consideration as it provided a simple
explanation for the discovery that pure chemicals react without remainders
only in certain weight (or volume) proportions (Dalton 1842). Strong support
for atoms came from the related discovery that all chemicals can be broken
down into (less than 100) ``elements,'' whose systematic relations were
successfully displayed in the periodic table (Meyer 1864, Mendel\'{e}ev
1869). Other support for the concept of atoms came from the gas laws and
their explanation with molecular kinetics (Maxwell 1859, Boltzmann 1868).
Nevertheless, in the phenomenological spirit of thermodynamics, 
opposition to the existence of atoms was still raised at the
turn of the century by extreme proponents (Mach 1896, Ostwald 1902) of the
greatest discovery in physics in the 1800s---the unifying principle of
conservation of energy (Mayer 1867). The atomic adversaries
(``energeticists'') conceded that atoms were useful thought patterns but not
reality. Their main argument was that mechanical laws, applied to atoms, 
could not account for irreversibility because of the independence of the 
equation of motion on the direction of time.
[In response to such doubts, and following Boltzmann's footsteps, Einstein, in his
doctoral thesis, put forth further arguments---molecular suspension---for
the existence of atoms (Einstein 1906).]

So it was in 1913, while on a post-doctoral fellowship in Manchester under
the tutelage of Ernest Rutherford, that Niels Bohr of Copenhagen proposed a
model of the atom that resolved the century-old mystery of spectral lines
(Bohr 1913a). Elaborating Rutherford's recent planetary concept of the atom,
Bohr combined Kepler's laws of celestial mechanics with the Planck-Einstein
quantization of energy (or alternatively, of angular momentum). The result
was a frequency formula [conceived three decades earlier with numerological
guesswork (Balmer 1885)] for the spectrum of the simplest
atom---hydrogen---that agreed with experiment to high precision. The same
method yielded a reasonable value for the ground-state energy of the next atom 
in the periodic table, helium, and promising results for some other atoms (Bohr
1913b). Extending the method to questions of the chemical bond, Bohr
obtained reasonable values for the binding and spacing of the hydrogen
molecule, $H_{2}$, but instability for the hydrogen molecule \textit{ion}, $%
H_{2}^{+}$ (Bohr 1913c).

Bohr's quantum theory found rapid acceptance not only because of its promise
to explain atomic structure, spectroscopy, and chemical bonding, but also
because of its mathematical simplicity---a back-of-the-envelope calculation
with high-school algebra. This was in contrast to Planck's derivation of
thermal radiation with demanding mathematics, applied to advanced
thermodynamics and electrodynamics. In further contrast to Planck's abstract
theory, the great conceptual appeal of the Bohr model lay in its
visualization with electron orbits and quantum leaps.

\section{BOHR MODEL}

The Bohr model of the hydrogen atom is like a miniature solar system where
the nucleus plays the role of the sun and the $H$ atom's single electron the
role of a planet. Although planet and electron are held in orbit by
different forces---gravitational and electric, respectively---these forces
have the same distance-dependence (inverse-square law). Thus both kinds of
orbit are ruled by Kepler's laws of planetary motion. There is, however, a
difference between gravity and electricity that poses a problem: According
to Maxwell's electrodynamics, any acceleration of electric charge (here,
centripetal acceleration of the orbiting electron) causes a radiative loss
of energy, so that the electron would spiral into the nucleus. No such
instability of the hydrogen atom is observed, though---in apparent
contradiction to the model. Bohr resolved the conflict with his postulate of
stationary states: Defying Maxwell's electrodynamics, the electron can move
in certain orbits---called ``quantum states''---without radiatively losing
energy. Bohr then imposed two restrictions on the multitude of possible
Kepler orbits: (i) A simplifying assumption---sufficient for energy
considerations---posits that only circular orbits are permitted. (ii) A
quantization condition requires that the angular momentum of the revolving
electron in the \textit{nth} orbit must equal $n$ times Planck's (reduced)
constant, 
\begin{equation}
L_{n}=n\hbar ,  \tag{1}
\end{equation}
where $\hbar $ $=h/2\pi $. The result are quantized circular electron orbits
whose radii depend on the square of the quantum number, 
\begin{equation}
r_{n}=n^{2}r_{B},  \tag{2}
\end{equation}
shown in Fig. 1. Here the Bohr radius $r_{B}=\hbar ^{2}/me^{2}$ $%
=0.529\times 10^{-10}m$ serves as a universal atomic length unit.
(Traditionally, the Bohr radius has been denoted as $a_{0}$. For the sake of
more ``user friendliness,'' the self-explanatory notation $r_{B}$ is used
here.) Note that the Bohr orbits are \textit{possible} orbits for the
revolving electron---the electron can, of course, only be in one such orbit
at a given time. 

\includegraphics[width=4in]{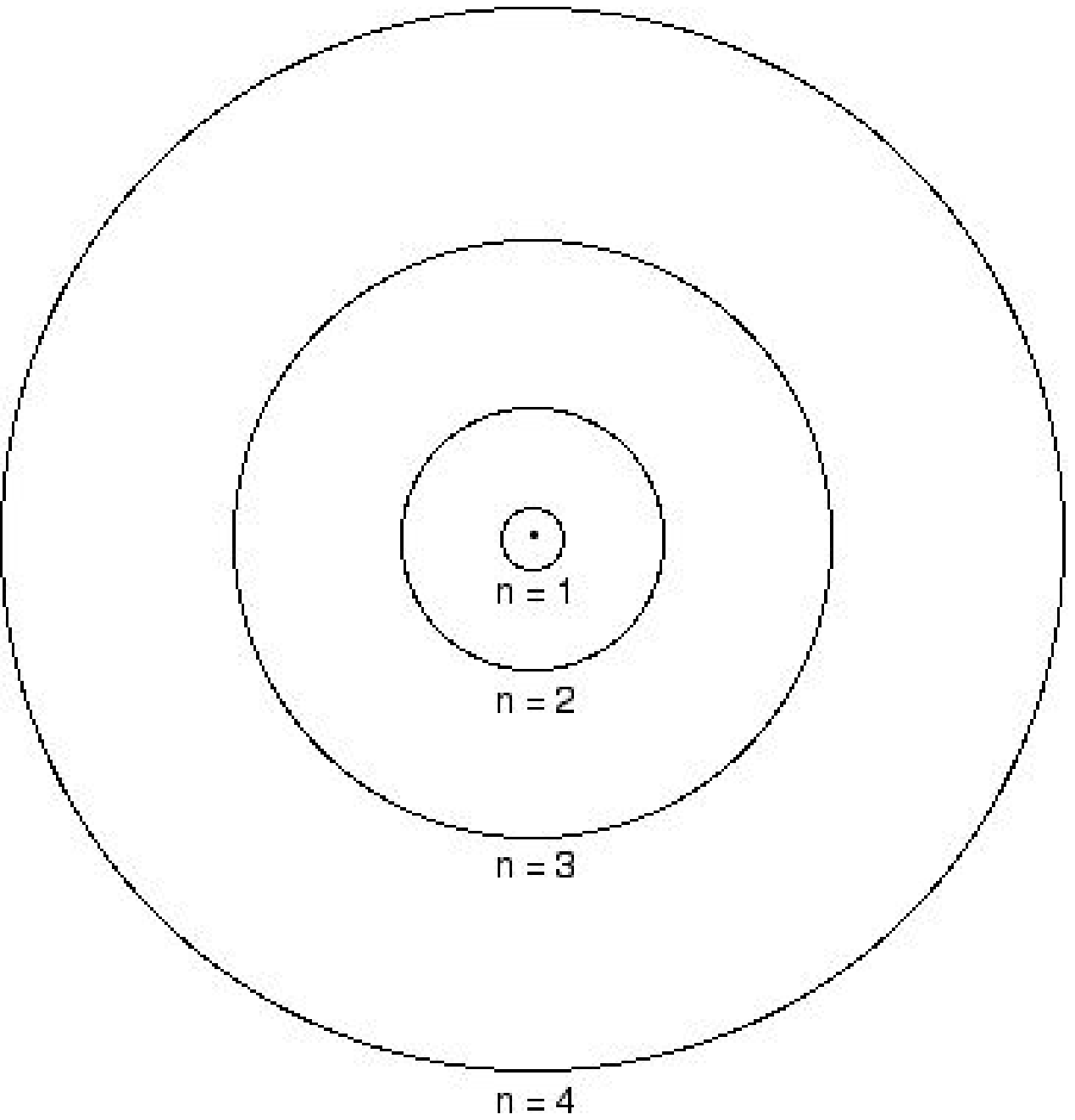}

\begin{quote}
Fig. 1. Bohr orbits of quantum number $n$.\bigskip
\end{quote}

\includegraphics[width=5in]{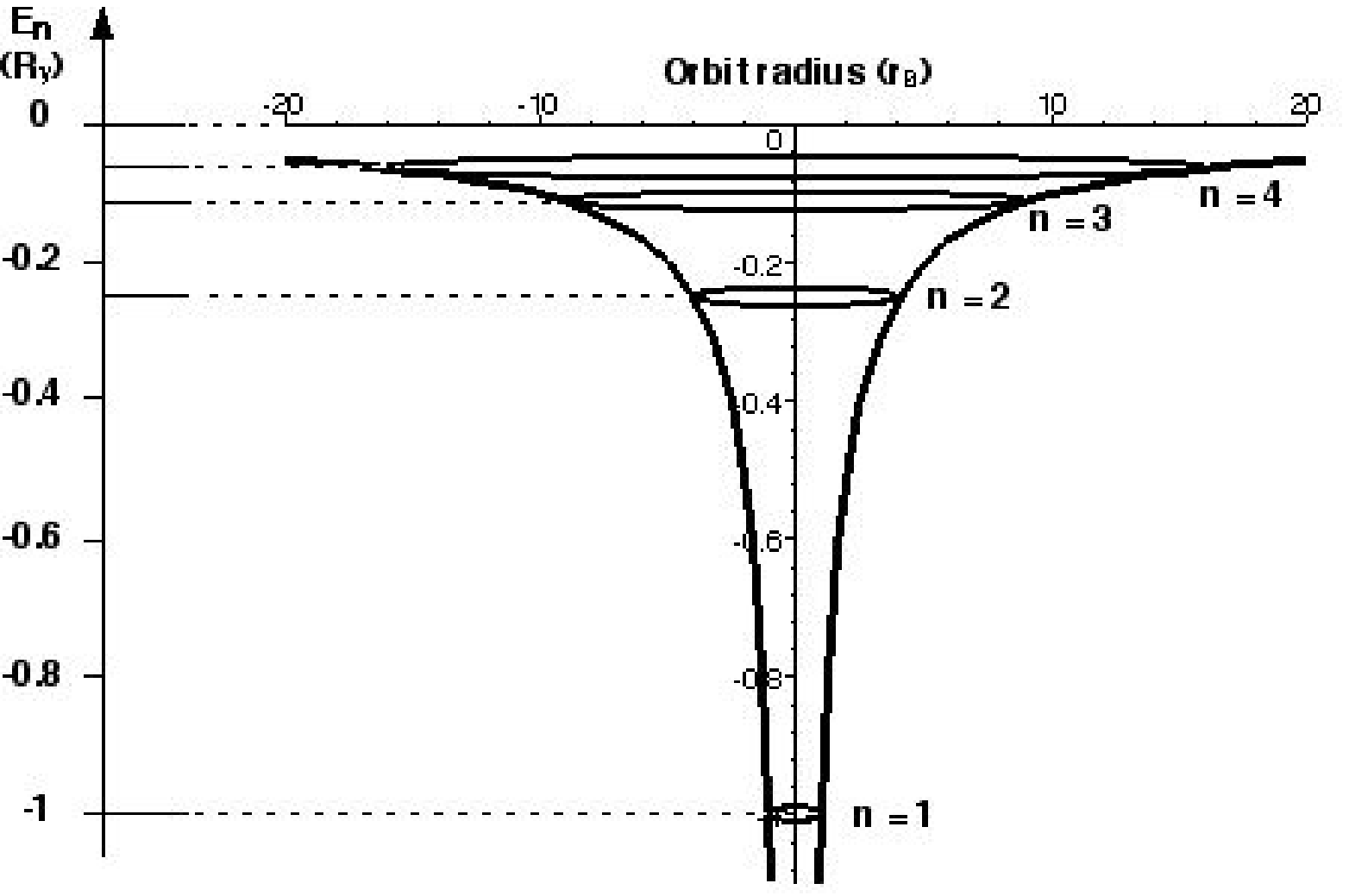}

\begin{quote}
Fig. 2. Coulomb funnel with inlaid Bohr orbits (center) and conventional energy-level diagram (left).\bigskip
\end{quote}

\noindent The quantized orbit energies are 
\begin{equation}
E_{n}=-\frac{1}{n^{2}}R_{y}.  \tag{3}
\end{equation}
The negative sign indicates that the orbits are bound and $R_{y}$ $%
=me^{4}/2\hbar ^{2}=13.6$ $eV$ is the Rydberg energy---an energy unit suited
to the atomic scale. Fitting, in the right part of Fig. 2, the Bohr orbits
into a Coulomb-potential funnel, $V(r)\propto 1/r$, visualizes the relation
with their energy levels, Eq. (3), which are commonly displayed by the
familiar level scheme in the left part of the figure.

The connection between orbit energies and spectral lines is achieved with
Bohr's postulate of quantum leaps: The electron can ``jump'' inward from an
outer orbit $n$ to an inner orbit $n^{\prime }$ by emission of a light
particle (photon) whose energy $\epsilon $\ is provided by the difference of
the orbits' energy,

\begin{equation}
\ E_{n}-E_{n^{\prime }}=\epsilon =hf_{nn^{\prime }},  \tag{4}
\end{equation}
which is, in turn, associated with the lines' frequencies $f_{nn^{\prime }}$
through the Planck-Einstein relation. The great success of the Bohr model,
achieved by combining Eqs. (4) and (3), is the Balmer formula, 
\begin{equation}
f_{nn^{\prime }}=\frac{R_{y}}{h}\left( \frac{1}{n^{\prime 2}}-\frac{1}{n^{2}}%
\right) \text{,}  \tag{5}
\end{equation}
in terms of quantum numbers and, via $R_{y}$, fundamental constants
(electron mass $m$ and charge $e$, Planck's constant $h$). Here $n$ denotes
the initial orbit and $n^{\prime }$ the final. A positive value of $%
f_{nn^{\prime }}$ signifies photon emission and a negative value photon
absorption. For a connection of the Balmer formula with experiment, one
converts the photon frequency to the corresponding wave length,
 $\lambda _{nn^{\prime}}=c/f_{nn^{\prime }}$, 
 according to their relation to the speed of light $c$. Constructive interference,
 $\lambda _{nn^{\prime}}=D$ $\sin \theta _{nn^{\prime }}$, in spectroscopy
 with a diffraction grating of groove separation $D$, permits observation of 
 spectral lines at diffraction angles $\theta_{nn^{\prime }}$.
 
A side result of the Bohr model is the  \textit{size} of the hydrogen atom, Eq. (2),
although no experiment was conceivable at the time to test this prediction.
The Bohr model is silent, though, on the relative brightness of the spectral
lines. Other predictions of the Bohr model are actually wrong: Results from
quantum mechanics, obtained over a decade later, prove that the (maximum)
angular momentum of the \textit{nth} quantum state is 
\begin{equation}
L_{n}=(n-1)h  \tag{6}
\end{equation}
instead of Eq. (1). Also disturbing, and jarring physical intuition, is the
Bohr model's spatial description of hydrogen as a ring atom instead of a
space-filling, spherical entity. Nevertheless, the Bohr model has been the
greatest coup in quantum physics. With bold assumptions it derived in a few,
strikingly simple steps a frequency formula that historically breached the
long-standing riddle of the spectral lines and provided a key to the
structure of the atom. It was therefore reasonable to expect that more
refined assumptions would resolve its short-comings and lead to further
progress. 

\includegraphics[width=3in]{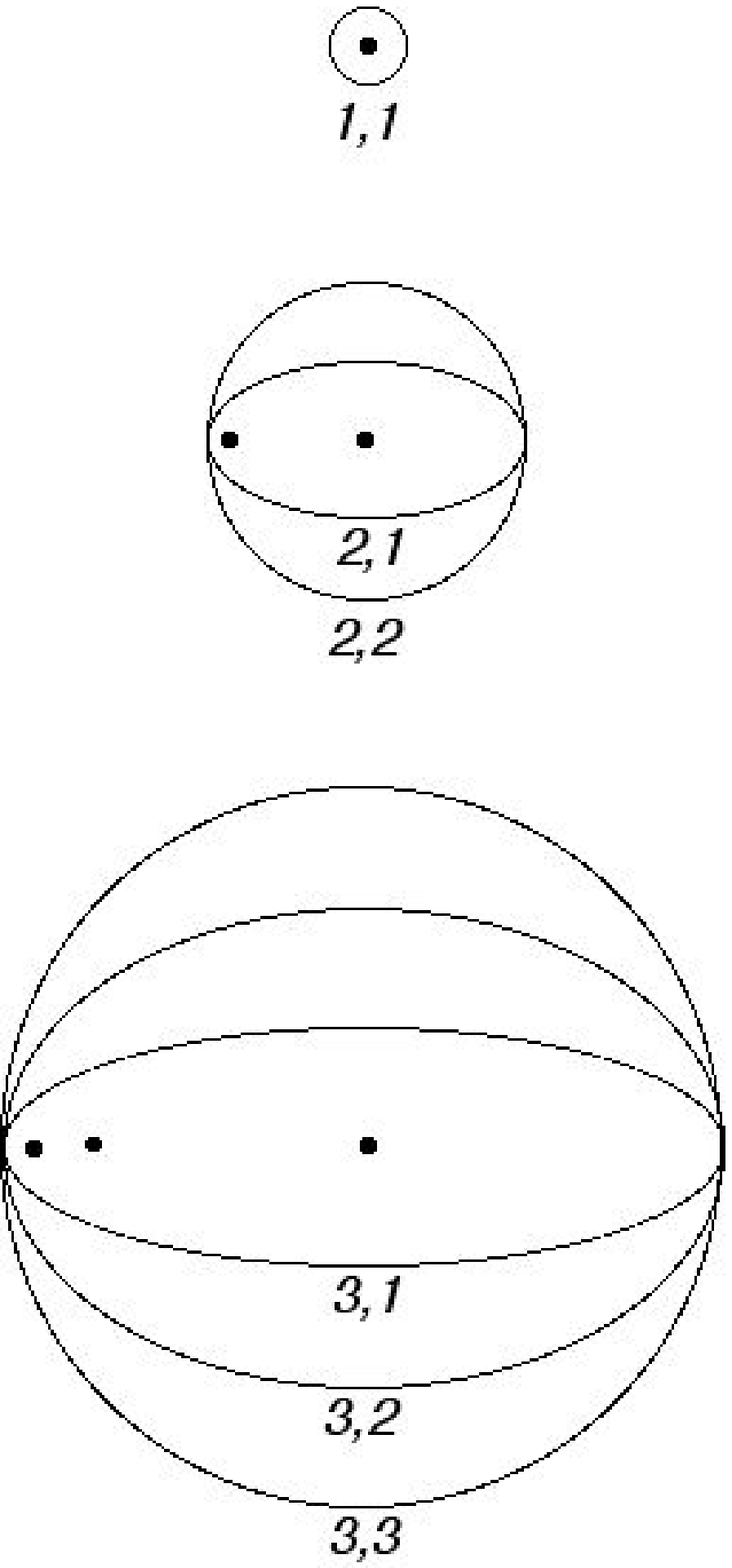}

\begin{quote}
Fig. 3. Coaxial $nl$ Sommerfeld ellipses with correspondingly shifted nuclei.
The quantum sum, $n$, characteristic for each orbit's \textit{energy}, determines 
its major axis.  By Kepler's 3rd law, the sorted orbits with the \textit{same} major axis 
have the same period and energy.
\end{quote}

\section{SOMMERFELD MODEL}

Such refinement occurred soon. Inspired by the fact that Planck's constant $h$ 
has the same dimension as the process quantity of action, $A$ = $\int pdq$, 
with $q$ being a generalized coordinate and $p$ the conjugate momentum, 
Arnold Sommerfeld, professor at the University of Munich, had previously 
sought a connection of quantization with the process quantity $A$---the quantity 
that is the backbone of the (least-) action-principle of mechanics. 
In this context he had coined for $h$ the term ``quantum of action'' 
(Sommerfeld 1911). Instead of Bohr's  \textit{ad hoc} quantization of 
angular momentum, Sommerfeld subsequently proposed that atomic 
process action should be quantized for each degree of freedom 
(Sommerfeld 1916). [The same quantization conditions had been proposed 
independently, albeit without application to atomic structure 
(Wilson 1915, Ishiwara 1915).] Lifting Bohr's special
assumption of circular orbits, Arnold Sommerfeld, professor at the
University of Munich, invoked the general case of Kepler motion: elliptical
orbits. He quantized separately the action of the electron's radial and
angular motion, 
\begin{equation}
\oint \ p_{r}dr=n_{r}h.  \tag{7}
\end{equation}

\noindent and 
\begin{equation}
\oint \ p_{\alpha }d\alpha =p_{\alpha }\oint d\alpha =2\pi p_{\alpha
}=n_{\alpha }h.  \tag{8}
\end{equation}

\noindent Here $\ p_{r}=m\ v_{r}$ is the radial momentum and $p_{\alpha }$
is the angular momentum---a constant of the motion, denoted $L_{n}$ in Eq.
(1). The sum of the radial and angular quantum numbers, 
\begin{equation}1
n_{r}+n_{\alpha }=n,  \tag{9}
\end{equation}
is the previous quantum number $n$ of the Bohr model, called (for short) 
\textit{``quantum sum''} by Sommerfeld. The angular quantum number was soon
relabeled as $k$ but will be denoted here as $n_{\alpha }\equiv l$ to
facilitate later comparisons with quantum mechanics. The electron orbits
that result from these assumptions are quantized ellipses, labeled $nl$,
with semimajor axis 
\begin{equation}
a_{nl}=n^{2}r_{B},\text{\quad }(n=1,2,...),  \tag{10}
\end{equation}

\noindent semiminor axis 
\begin{equation}
b_{nl}=nlr_{B},\quad (l=1,2,...,n),  \tag{11}
\end{equation}

\noindent and orbit energy 
\begin{equation}
E_{nl}=-\frac{1}{n^{2}}R_{y}.  \tag{12}
\end{equation}

Note in Eqs. (10, 12) that both the orbit's semimajor axis and energy depend
only on the quantum sum $n$ and agree with the radius and energy of the 
\textit{nth} Bohr orbit, Eqs. (2, 3), respectively. Accordingly, for a given
quantum sum $n$, all $nl$ Sommerfeld ellipses $(l=1,2,...,n)$ have the same
length of major axis and the same energy. They include the circular Bohr
orbits as a special cases, $l=n$. The lesser the angular quantum number $l$,
the sleeker the ellipse. Orbits with the same length of major axis (same
energy) but of different shape (different angular momentum) are called
``degenerate.'' By Eqs. (10 - 12), the $nl$ Sommerfeld orbits are $n$-fold
degenerate. The arrangement, in Fig. 3, of Sommerfeld ellipses of the same
quantum sum $n$ by groups with coinciding major axes shows such 
degenerated quantum states (Sommerfeld 1931). [The elliptical orbits' 
energy levels, Eq. (12), can, of course, also be visualized by the fit of their 
major axes into the Coulomb funnel of Fig. 2.]

The \textit{minor} axis of a Sommerfeld ellipse depends on both the quantum sum $n$
and the angular quantum number $l$, Eq. (11). In contrast, an ellipse's
semilatus rectum, 
\begin{equation}
\Lambda _{nl}=l^{2}r_{B}\quad (n=1,2,...)  \tag{13}
\end{equation}
---the distance from the focus to the perimeter, parallel to the minor
axis---depends exclusively on the angular quantum number $l$ which in turn
determines the orbit's angular momentum $L_{nl}=l\hbar $, Eq. (8).
Geometrically, the semilatus rectum $\Lambda _{l}$ $\equiv $ $\Lambda _{nl}$
is the radius of curvature of an ellipse at its major vertex. Thus the
angular momentum of an $nl$ Sommerfeld ellipse is represented by its vertex
curvature (Mallow 1999) as the comparison in Fig. 4 shows.

Missing among the Sommerfeld orbits is one more solution of Eq. (11), $l=0$
---the straight-line ellipse where the electron collides with the nucleus.
Sommerfeld deliberately excluded this case as \textit{``unphysical.''}

\includegraphics[width=3in]{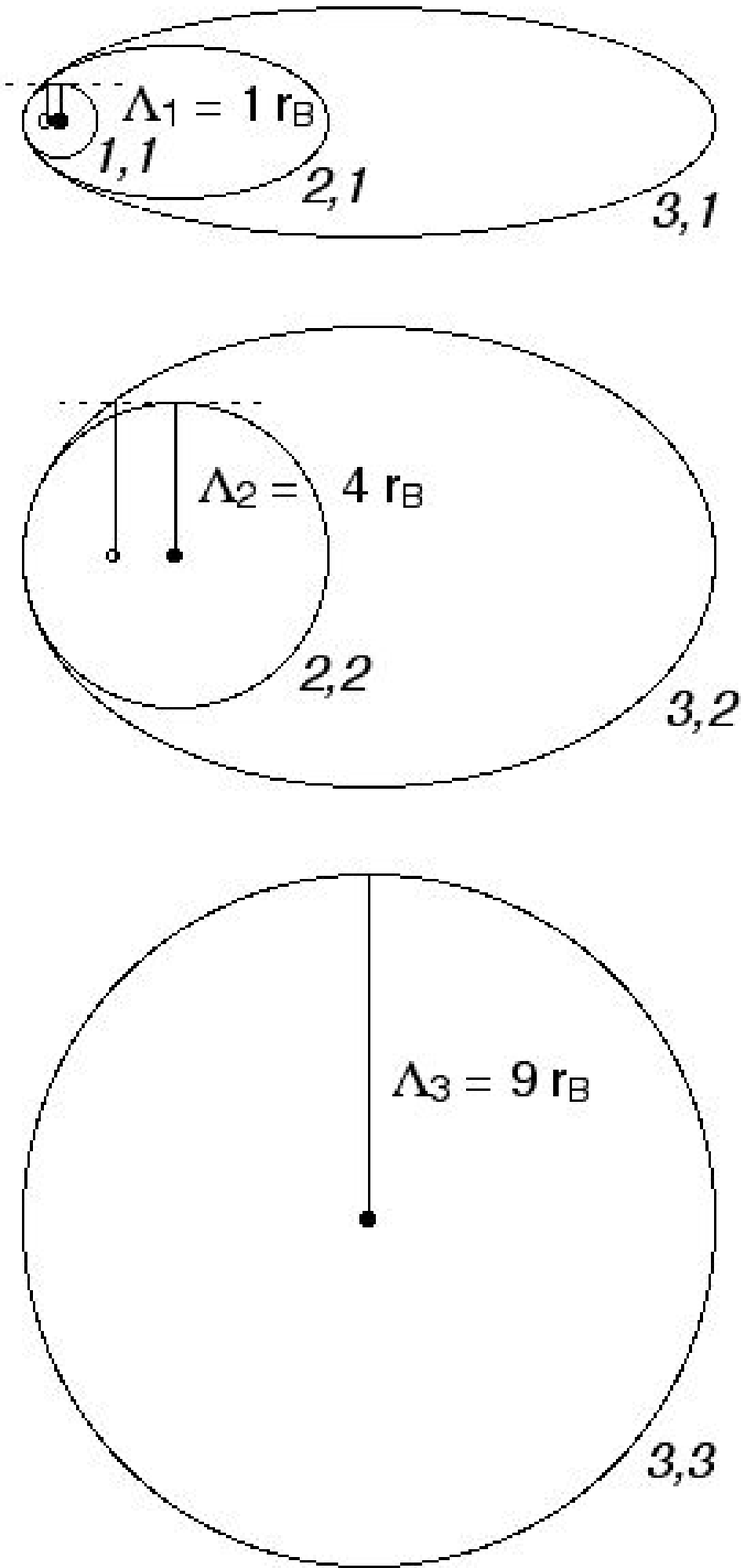}

\begin{quote}
Fig. 4. Coverticil $nl$ Sommerfeld ellipses with correspondingly shifted nuclei. 
The angular quantum number, $l$, characteristic for each orbit's 
\textit{angular momentum}, determines its \textit{semilatus rectum} $\Lambda _{l}$.  
Geometrically, $\Lambda _{l}$ is the radius of curvature of an ellipse at its vertex.  
Therefore, all ellipses with the \textit{same} $l$ have the same \textit{curvature}.
\end{quote}

A description of motion in three-dimensional space requires three
coordinates---Cartesian $(x,y,z)$ or, better suited for atoms, spherical $%
(r,\theta ,\varphi )$. The use of planar polar coordinates $(r,\alpha )$ in
Eqs. (7, 8) amounts to a two-dimensional quantization, whose coplanar
ellipses serve as a conceptual stepping stone for Sommerfeld's quantization
of motion in three ``degrees of freedom'' of space: 
\begin{equation}
\oint \ p_{r}dr=n_{r}h,  \tag{14}
\end{equation}
\begin{equation}
\oint \ p_{\theta }d\theta =n_{\theta }h,  \tag{15}
\end{equation}
and 
\begin{equation}
\oint \ p_{\varphi }d\varphi =\ p_{\varphi }\oint \ d\varphi =2\pi\
p_{\varphi }=n_{\varphi }h.  \tag{16}
\end{equation}
The previous angular action integral, Eq. (8), splits now into latitudinal
and equatorial action integrals, Eqs. (15, 16), such that the angular
quantum sum is 
\begin{equation}
n_{\theta }+n_{\varphi }=n_{\alpha }\equiv l.  \tag{17}
\end{equation}

Both the angular momentum $p_{\alpha }$, Eq. (8), and its the axial
component, $(p_{\alpha })_{z}\equiv p_{\varphi }$, Eq. (16), are constants
of the motion, which justifies their factoring out of the action integrals
in these equations. The quantum number of the latter is traditionally called
``magnetic'' and relabeled as $n_{\varphi }$ $\equiv $ $m_{l}$. A more
descriptive term for $m_{l}$ would be \textit{``tilt} quantum number'' which
we wish to use henceforth. The orbits that result from the 3D quantization,
Eqs. (14-16), are the same Sommerfeld ellipses as from the 2D quantization,
Eqs. (7, 8), but now oriented at quantized tilt angles, 
\begin{equation}
\tau _{m_{l}}=\arccos (m_{l}/l),\qquad (m_{l}=0,\pm 1,\pm 2,...,\pm l), 
\tag{18}
\end{equation}
(``space quantization'') with respect to a reference plane---commonly
equatorial---depending on the possible values of $m_{l}$ . Accounting for
all orientations, a given $nl$ Sommerfeld ellipse has then a $(2l+1)$-fold
orientational multiplicity. An example of some of the possible tilts of an $%
nl=(n,3)$ Sommerfeld orbit (here, only those \textit{above} the $xy$
reference plane, $m_{l}=0,+1,+2,+3$) is shown in the upper part of Fig. 5.
The concept of space quantization is visualized in the lower part of Fig. 5
by the projected areas of the variously tilted $nl$ ellipse onto the
equatorial plane. The sizes of the projected areas proportionally represent
the equatorial action, Eq. (16). [Both the tilted $nl$ ellipse and
each projection have the same semimajor axis, $a_{nl}$, Eq. (10). The semi-

\includegraphics[width=5in]{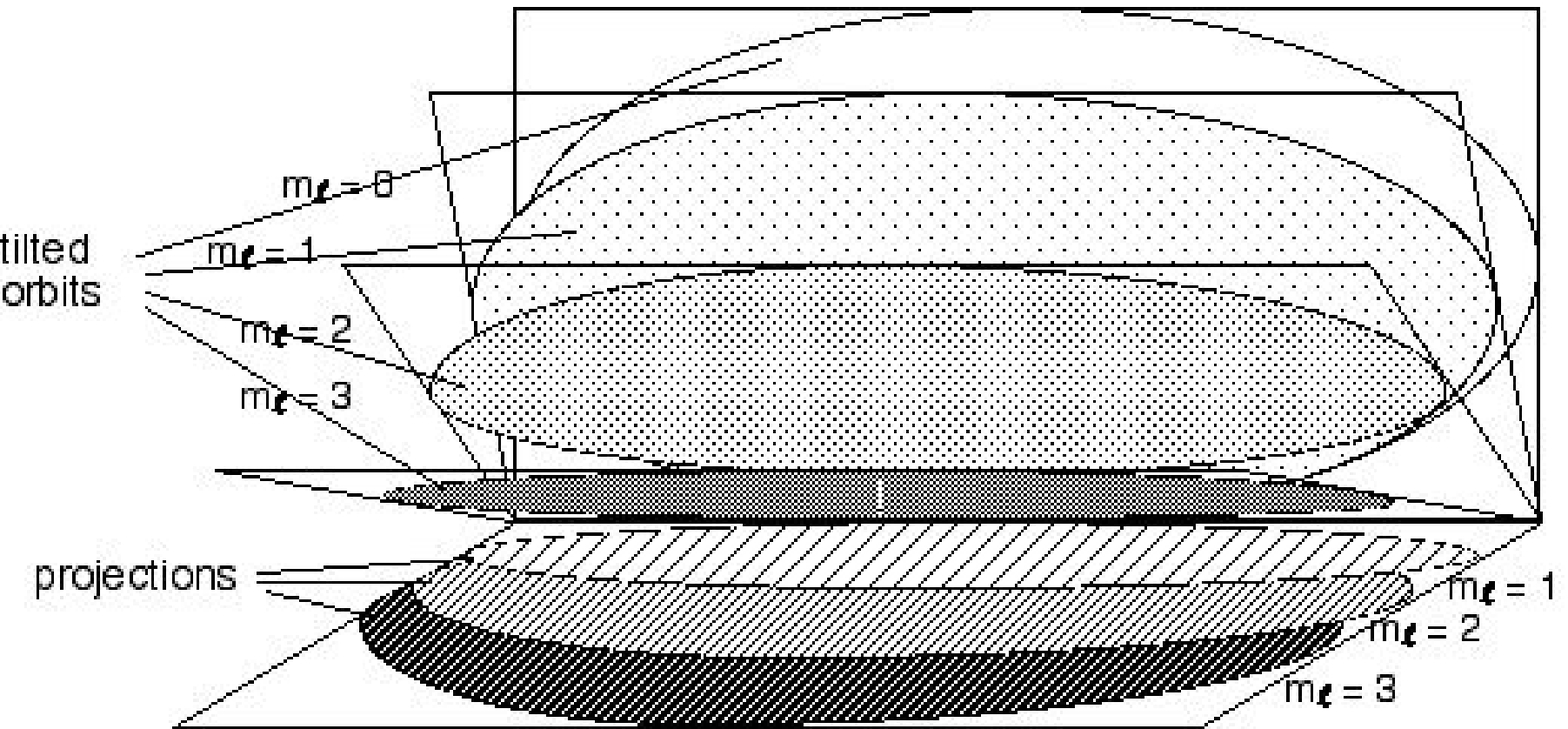}

\begin{quote}
Fig. 5. Orbital ellipse of angular quantum number $l=3$ in different spatial orientations and their projections onto the equatorial plane.
\end{quote}

\noindent  minor axis of the tilted ellipse is $b_{nl}$, Eq. (11), and that of the projected
ellipse is $b_{nm_{l}}=nm_{l}r_{B}$. By the area formula, the ratio of the
projected area to the area of the entire, tilted ellipse is $\pi
nm_{l}r_{B}\times $\ $n^{2}r_{B}/\pi nlr_{B}\times n^{2}r_{B}=m_{l}/l$.] For
the oriented $nl$ ellipse only such tilt angles $\tau _{m_{l}}$ are allowed
which result in projected area sizes in the ratio of Eq. (18).

The hierarchy of tilted ellipses, with their ordering by quantum sum $n$
(length of major axis), angular quantum number $l=1,2,...,n$ (vertex
curvature), and tilt quantum number $m_{l}=0,\pm 1,\pm 2,...,\pm l$ (space
quantization), has become invaluable for an understanding of the structure
of \textit{all} atoms---not only hydrogen---and the classification of
spectroscopy, particularly in the presence of a magnetic field (Zeeman
effect). Because of its electric charge, $-e$, an electron's motion along a
tilted orbit amounts to a closed-loop current and thus to an electromagnet.
The field energy of such orbit-magnets in an external magnetic field $B$
lifts the orbit degeneracy by distinguishing the tilted orbits energetically
(splitting of spectral lines).

Even without a magnetic field, a very minute splitting of hydrogen spectral lines
had been observed already a generation before the old quantum theory
(Michelson 1891). One motivation for Sommerfeld's extension of the Bohr
model was to account for this \textit{fine structure}. Since Sommerfeld's
procedure, as shown above, yields the same orbit energies as Bohr's, Eqs.
(12, 3), it fails to explain the fine structure. Therefore Sommerfeld
refined his previous quantization scheme by including the electron's
relativistic mass dependence, $m=m_{0}/\sqrt{1-(v/c)^{2}}$ , where $m_{0}$
is the electron's rest mass. The result is, besides the strong dependence of
the orbit energy on the quantum sum $n$, a slight dependence on the angular
quantum number $l=1,2,...,n$, 
\begin{equation}
E_{nl}=m_{0}c^{2}\left\{ \left[ 1+\frac{\alpha ^{2}}{\left( n-l+\sqrt{
l^{2}-\alpha ^{2}}\right) ^{2}}\right] ^{-\frac{1}{2}}-1\right\} ,  \tag{19}
\end{equation}

\noindent due to near cancellation of $-l$ in Eq. (19) by the radical following it. 
Here $m_{0}c^{2}$ is the electron's rest energy and $\alpha
=e^{2}/\hbar c\cong 1/137$ is the fine-structure constant. A power-series
expansion to order $\alpha ^{2}$,

\begin{equation}
E_{nl}\cong -\left[ \frac{1}{n^{2}}+\frac{\alpha ^{2}}{n^{4}}\left( \frac{n}{%
l}-\frac{3}{4}\right) \right] R_{y},\quad (l=1,2,...,n),  \tag{20}
\end{equation}

\noindent contains, by the first term in the square bracket, the previous,
non-relativistic orbit energy, Eq. (12), and the additional relativistic
correction.

Sommerfeld's fine-structure formula was spectroscopically confirmed for $H$
and (in its more general form, not shown here) for hydrogenic ions ($He^{+}$, 
$Li^{2+}$, ...) to the highest precision (Paschen 1916)---the greatest
triumph of the old quantum theory. Over a decade later, after the electron
spin had been discovered (Uhlenbeck, Goudsmit 1925), relativistic quantum 
\textit{mechanics} (Dirac 1928) obtained the fine-structure formula (Gordon
1928) along a very different route, stated in slightly modified terms, 
\begin{equation}
E_{nl}=m_{0}c^{2}\left\{ \left[ 1+\frac{\alpha ^{2}}{\left( n-l+\sqrt{%
\lambda ^{2}-\alpha ^{2}}\right) ^{2}}\right] ^{-\frac{1}{2}}-1\right\} . 
\tag{21}
\end{equation}

\noindent The parameter $\lambda $ depends on the electron's \textit{total}
angular quantum number $j=l\pm \frac{1}{2}$, composed of its \textit{orbital 
}angular quantum number $l$ and spin quantum number $s=\frac{1}{2}$ such
that $\lambda =l+1$ for $j=l+\frac{1}{2}$ and $\lambda =l$ for $j=l-\frac{1}{%
2}$. Expanded to order $\alpha ^{2}$, this formula becomes 
\begin{equation}
E_{nl}\cong -\left[ \frac{1}{n^{2}}+\frac{\alpha ^{2}}{n^{4}}\left( \frac{n}{%
j+\frac{1}{2}}-\frac{3}{4}\right) \right] R_{y},\quad \left( j=\frac{1}{2},%
\frac{3}{2},\text{ }...\right) ,  \tag{22}
\end{equation}

\noindent and yields the same energies as Sommerfeld's formula, Eq. (20),
(although with an altered ordering of quantum numbers).

\section{HELIUM ATOM AND HYDROGEN MOLECULE ION}

Sommerfeld-like $(n,l,m_{l})$ orbits, populated with electrons from the
inside out, conceptually provide the backbone of the atomic \textit{Aufbau}
(building-up) principle (Bohr 1923). [It would later explain, when combined
with the exclusion principle (Pauli 1925b), the periodic table of the
elements.] Atoms other than hydrogen constitute a many-body problem, defying
closed-form solutions. Accordingly, one has to resort to approximation
methods, essentially by reducing the multi-electron situation to an
effective one-electron problem through shielding by the atomic core. In his
initial model of the \textit{helium} atom, Bohr assumed that the atom's two
electrons circulate, diametrically opposite, in a quantized ring orbit about
the nucleus (Bohr 1913b), illustrated in Fig. 6. His value for the $He$
ground-state energy was surprisingly close (within 5\%) to experiment. Much
disappointment followed therefore when subsequent refinements with tools of
Sommerfeld's orbit scheme (Born and Pauli 1922, Born and Heisenberg 1923)
and tremendous calculational efforts (Van Vleck 1922, Kramers 1923) failed
to yield better results.\bigskip

\includegraphics[width=2in]{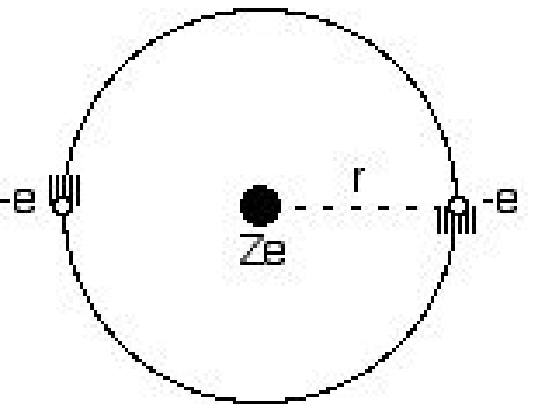}

\begin{quote}
Fig. 6. Bohr's historic "two-seat-roundabout" model of the $He$ atom.\bigskip
\end{quote}

Similar difficulties plagued orbital calculations of other atoms. The
assignment of an angular momentum to the atom core (Land\'{e} 1921), and its
association with an ``inner'' quantum number, was eventually proved
inconsistent (Pauli 1925a). Understanding the anomalous Zeeman effect
remained an insurmountable barrier for the old quantum theory---desperate
efforts even involved attempts with half-integer quantum numbers (vindicated
after the discovery of spin).

The hydrogen molecule ion, $H_{2}^{+}$, was another proving ground for
quantum theory. A one-electron problem, subject to attraction to two nuclei,
it should be expected to be solvable. In exploration, Bohr (1913c) had
modeled $H_{2}^{+}$ with a quantized ring orbit in the bisectorial plane
between the nuclei but deemed it unstable by mechanical considerations.
Sommerfeld calculated $-0.88R_{y}$ for the ground-state energy of this ring
orbit which renders it \textit{energetically} unstable. (If the energy of
the $H_{2}^{+}$ ground state is higher than that of the $H$ atom, $-1R_{y}$,
then the molecule ion is unstable and decays into a free $H$ atom and a
far-away proton.) Thereupon, in 1920, Sommerfeld put his most gifted
student, the prodigy Wolfgang Pauli, to the task for a more detailed
analysis of the $H_{2}^{+}$ problem. Drawing on established methods of
celestial mechanics (Jacobi 1866), Pauli classified possible electron
orbits: four planar orbits, illustrated in Fig. 7---(a) ``pendulum orbit''
in the median plane perpendicular to the nuclear axis, (b) ``satellite
orbit,'' sweeping the region about one nucleus, (c) ``lemniscate orbit,''
densely filling the area of an ellipse with both nuclei as foci, (d)
``planetary orbit'' over the interval area between two ellipses around both
nuclei---and two groups of spatial orbits (not shown in Fig 7), symmetric or
asymmetric to the median plane (Pauli 1922). The pendulum orbits, $(a)$, are
unstable against displacement perpendicular to their oscillation. Planetary
orbits, $(d)$, fail to satisfy a time-average equilibrium condition.
Remarkably, Pauli discarded the satellite and lemniscate orbits, $(b, c)$,
because they lead arbitrarily close to a nucleus. For justification, Pauli
referred to a (unstated) rule by Bohr, according to which such orbits would
not occur ``in reality.''

With all planar orbits excluded, Pauli analyzed the symmetric spatial
orbits. (He postponed the asymmetric case because of excessive numerical
effort.) Appropriate to the molecule's symmetry, the orbits are expressed in
spheroidal coordinates, $\xi =(r_{1}+r_{2})/R,\eta =(r_{1}-r_{2})/R$ and $%
\varphi $. Here $r_{1}$ and $r_{2}$ are the electron's distance from the
left and right nucleus, $R$ is the nuclear separation, and $\varphi $ is the
equatorial angle in the median plane. Using these coordinates and
corresponding momenta, the quantization conditions are analogues to Eqs.
(14-16),

\begin{equation}
\oint \ p_{j}dj=n_{j}h,\quad (j=\xi ,\eta ,\varphi ).  \tag{23}
\end{equation}

Pauli found that the orbit characterized by the quantum numbers $(n_{\xi
},n_{\eta },n_{\varphi })=(0,1,1)$ possessed the lowest energy. Accordingly
he identified it as the

\includegraphics[width=5in]{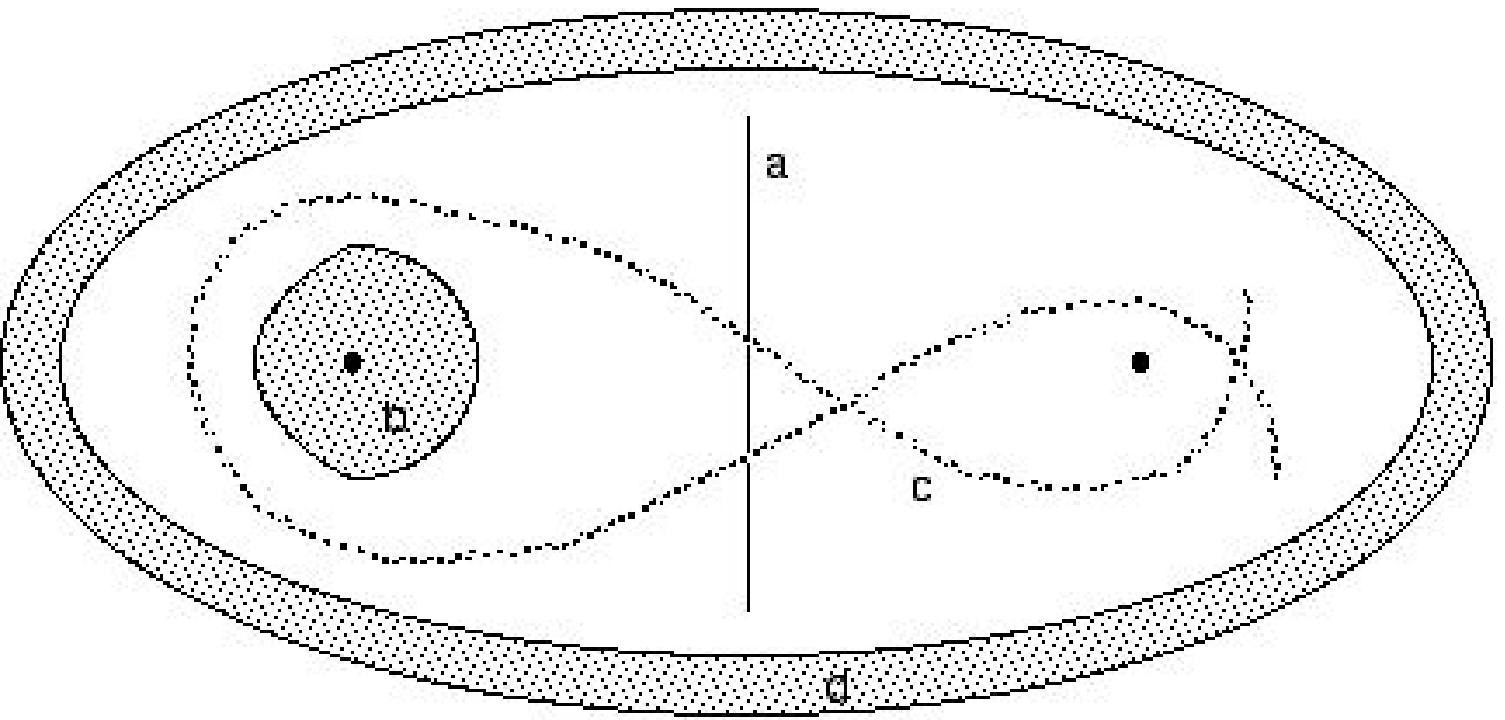}

\begin{quote}
Fig. 7. Pauli's classification of planar electron orbits in the hydrogen molecule ion, $H_{2}^{+}$: (a) pendulum orbit, (b) satellite orbit, (c) lemniscate orbit, (d) planetary orbit\bigskip
\end{quote}

\noindent molecule ion's ground state. Note that the $(0,1,1)$
quantization requires \textit{two} quanta of action---in contrast to $1h$
for the ground state of the $H$ atom. The calculated energy of the $%
H_{2}^{+} $ ground state came out as $-0.52R_{y}$. [The same result was
obtained independently by K. F. Niessen of Holland (Niessen 1923).] As a
consequence the hydrogen molecule ion would be unstable. In contrast,
contemporary experiments indicated stability of $H_{2}^{+}$, but some
uncertainty remained. A few years later, though, experiments clearly
confirmed that the hydrogen molecule ion was stable, with a ground-state
energy of $-1.20R_{y}$.

The $H_{2}^{+}$ problem presents a paradigm for the chemical bond---two
nuclei held together by an electron (Gutzwiller 1990). Pauli's failure to
find a stable ground-state of $H_{2}^{+}$ was a serious blow to the old
quantum theory. Something seemed amiss.

\section{FALL OF THE OLD QUANTUM THEORY}

The final straw that broke the old quantum theory was the hydrogen atom in
crossed electric and magnetic fields. The spectral response of the atom to
only one such field is known as the Stark effect or the Zeeman effect,
respectively. In his theoretical treatment of the crossed-fields problem,
Pauli invoked Ehrenfest's adiabatic principle whereby quantization persists
under slow variation of a parameter (Ehrenfest 1916). In this way Pauli
could restate the situation such that the electron would collide with the
nucleus---Sommerfeld's ``unphysical'' situation---and show that the old
quantum theory is \textit{contradictory} (Pauli 1926).

What exactly was the root of the old quantum theory's failures? Pauli,
Heisenberg and Born suspected that is was the \textit{orbit paradigm} that
caused the difficulties. Subsequent mathematical explorations of more
abstract nature (differential operator techniques) by Born and Heisenberg
led to the discovery of matrix mechanics, based solely on quantities that
can be observed in principle (Heisenberg 1925, Born et al. 1925). The
successful solution of the previous difficulties with the $H$ atom with the
new method---called, for distinction, quantum \textit{mechanics}---seemed to
confirm the inappropriateness of the orbit concept (Pauli 1926). This notion
acquired formal support with the discovery of the uncertainty principle
(Heisenberg 1927). Henceforth the old quantum theory of Bohr and Sommerfeld
was abandoned from serious calculations. Only occasionally was it employed,
with many caveats and much warning, for introductory purposes (Bohr model,
visual appeal of Sommerfeld orbits. [Ironically---and confusing generations
of students---some texts and handbooks labeled the Sommerfeld orbits
faultily (Jaworski and Detlaf 1972). They erroneously assigned the angular
quantum number $l=0$ to the circular (Bohr) orbits, and orbits of increasing 
$l$ to \textit{in}creasing eccentricity. It is not clear what caused such
confusion---perhaps mistaken association of Sommerfeld's circular
ground-state orbit with the spherical symmetry of the ground state's
quantum-mechanical description? Or, paradoxically, a formal application of
unvisualizable, quantum-mechanical rules to a quantum visualization by
orbits, starting with the first Bohr orbit as ground state?] Quantum
mechanics, while suited to calculate atomic properties to highest precision,
left little room for visualization---a serious curtailment of physical
intuition. [Quantum mechanics' few visualization tools are graphs of radial
or angular wave functions $\psi $ (by themselves entities of no or vague
interpretation), squared wave functions $|\psi |^{2}$ with their probability
interpretation (Born 1926), and spatial representations of orbitals'
electron-density to serve as indicators for chemical bonding.]

\section{COULOMB OSCILLATOR}

Eighty years after its demise the old quantum theory experienced a
resurrection when a fatal flaw in the Sommerfeld model was recognized: the
exclusion of the ``unphysical'' electron collision with the
nucleus---Sommerfeld's dictum $l\neq 0$, Eq. (11). If instead the electron
is permitted to swing through the nucleus (``Coulomb oscillator,'' $l=0$),
the correct hierarchy of orbital angular momenta in the hydrogen atom is obtained (Bucher 2006).
(How I came to that revelation, I tell in the appendix.) Likewise, when the
electron of $H_{2}^{+}$ swings through both nuclei with an action of $1h$,
stability of the molecule ion is achieved. By excluding this case, both
Pauli and Niessen had missed that crucial result. The unstable orbit they found, 
 with action of $2h$,

\includegraphics[width=5in]{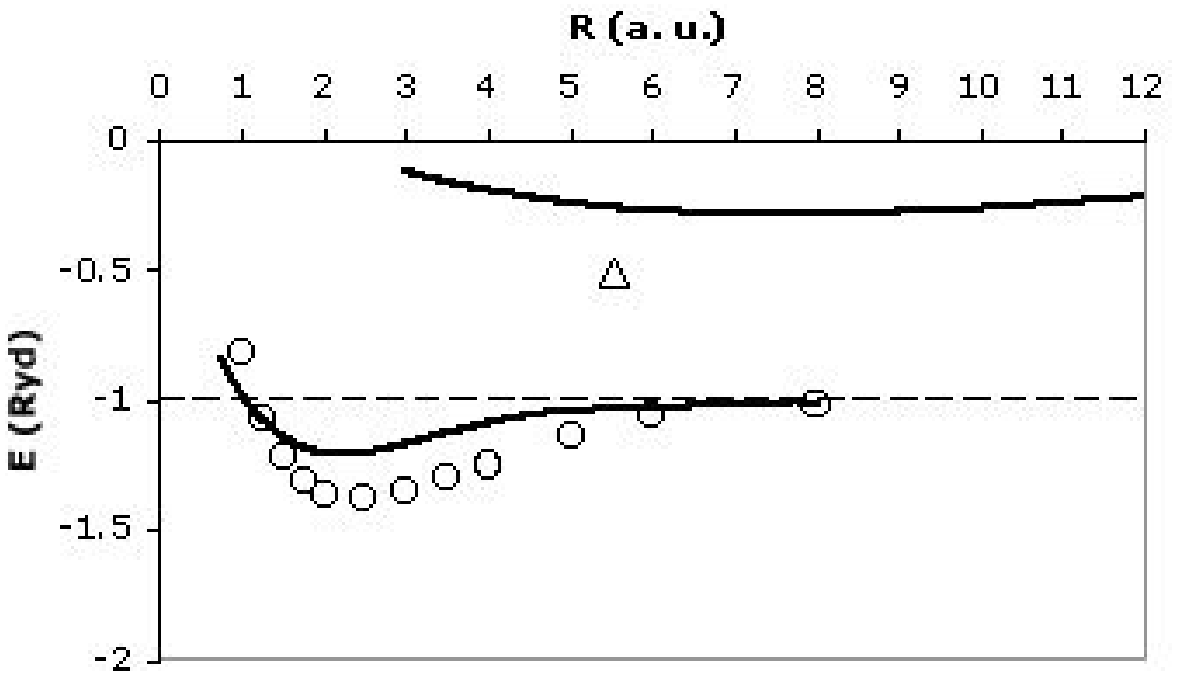}

\begin{quote}
Fig. 8.  Binding energy $E$ of the hydrogen molecule ion $H_{2}^{+}$ \textit{vs} nuclear separation $R$:  Exact, quantum mechanical solution (curve) and Coulomb-oscillator calculation (circles).  The bottom curve is the $1s$$\Sigma _{g}$ ground state, the top curve the $2p$$\Pi _{u}$ excited state.  The triangle shows Pauli and NiessenÕs historical calculation (at equilibrium), misinterpreted as the (unstable) ground state.  Data beneath (above) the horizontal dashed line---the binding energy of a free $H$ atom---indicate stability (instability) of the molecule ion.
\end{quote}

\includegraphics[width=5in]{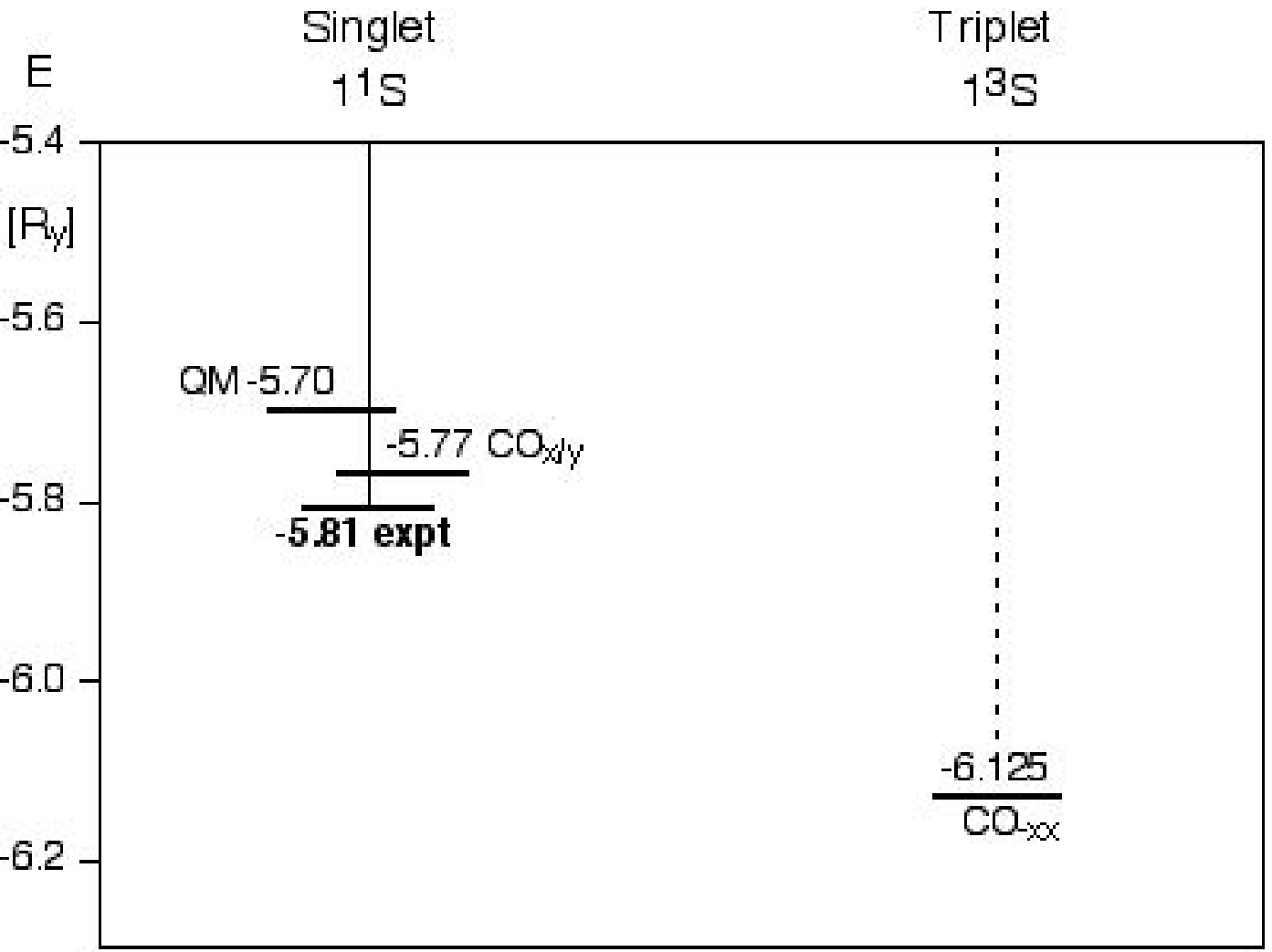}

\begin{quote}
Fig. 9.  Experimental energy level (in $R_{y}$) of the helium ground state and calculated values with variational quantum mechanics (QM) and with Coulomb oscillations (CO) in contralinear $(-xx)$ and cross-delayed $(x/y)$ modes.
\end{quote}

\noindent can today be identified as the approximate
solution of an excited state of $H_{2}^{+}$, rather than their mistaken
ground-state allocation (see Fig. 8). The small deviation of the
Coulomb-oscillator calculation of $H_{2}^{+}$ from the exact solution, Fig.
8, is caused by the former's neglect of the electron's wave nature (Bucher
2006).

Applied to the \textit{helium} atom (Bucher 2007), synchronized
cross-oscillations of \textit{two} electrons through its nucleus give the
energy of the $1^{1}S$ ground state of $He$, in close agreement with
experiment (see Fig. 9). In contrast, contralinear oscillations give the
same energy as Bohr's historical two-seat roundabout model, also shown in
Fig. 9, and are interpreted as the forbidden $1^{3}S$ triplet ground
state---a result beyond the reach of quantum mechanics. 

Finally, Pauli's
discovery in his crossed-fields analysis of the $H$ atom---that the electron
would collide with the nucleus---proves to be no contradiction but\textit{\
reality}.

\includegraphics[width=5in]{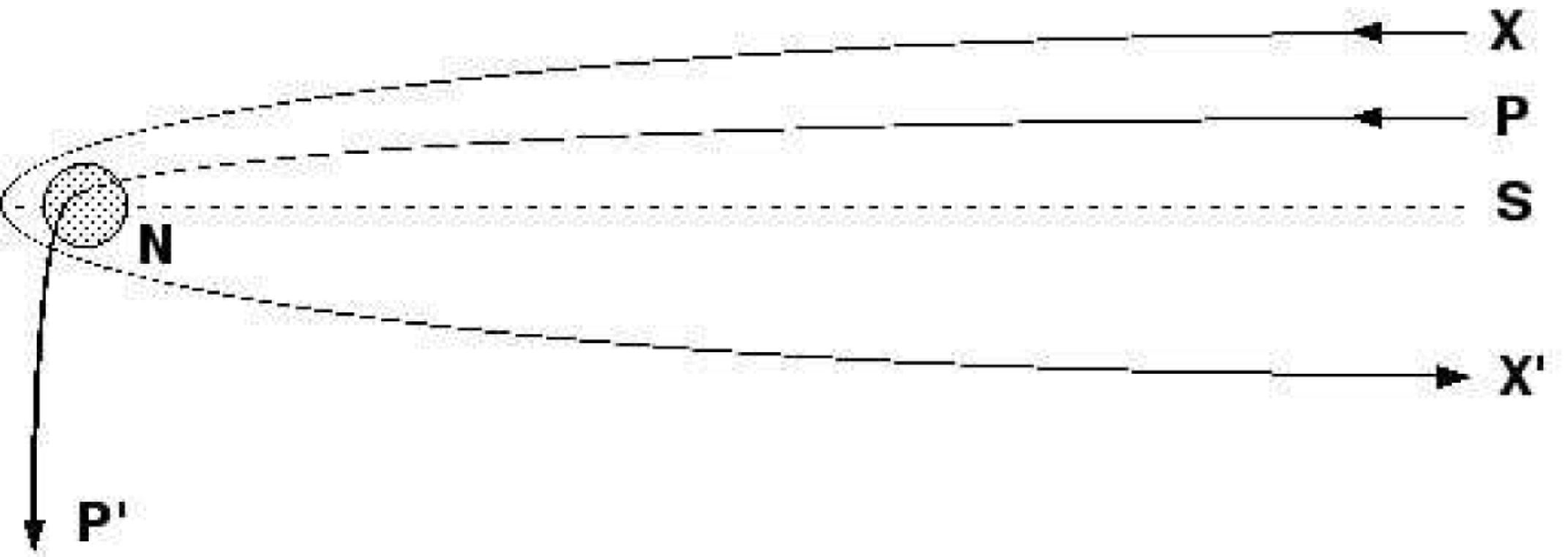}

\begin{quote}
Fig. 10.  Partial trajectory of an extranuclear, elliptical orbit $XX'$  and of a penetrating orbit $PP'$  through a finite-size nucleus $N$.  The dotted line $S$ shows the major symmetry axis of $XX'$. \end{quote}

Obviously, the case of a planet colliding with the sun---and reappearing
unscathed on the opposite side---is missing in the classical treatises of
celestial mechanics (Jacobi 1866, Poincar\'{e} 1905, Charlier 1907). Still,
it is surprising how matter-of-factly Sommerfeld dismissed this scenario on
the atomic scale as ``unphysical.'' Likewise, the possibility of a
rectilinear electron oscillation through the nuclei is never considered in
Pauli's treatment of $H_{2}^{+}$, in contrast to his systematic analysis of
all other orbits.

A subtle, conceptual distinction arises for the rectilinear electron orbits,
depending on the assumed size of the nucleus---finite or infinitely small.
Closed electron orbits are always elliptical if they pass \textit{outside}
the nucleus. It is the entire nuclear charge, $+Ze$, that causes the
centripetal acceleration. In the (temporarily assumed) continuous 
limit of maximal elliptical eccentricity, $%
l\rightarrow 0$, the orbit becomes a straight line of (major-axis) length $%
2n^{2}r_{B}$, Eqs. (10, 11), with its nuclear and empty focus at either
end---provided the nucleus is a \textit{point} charge! The centripetal
acceleration of the electron then occurs infinitely close to the point
nucleus such that the electron motion appears as a repulsion at the (point)
nucleus---``Coulomb repellor.'' A different situation is at hand, though,
when the electron penetrates a \textit{finite-size} nucleus. In this case
only \textit{part} of the nuclear charge affects the centripetal
acceleration so that no closure of (genuine) elliptical orbits can occur
(see Fig. 10). In the case of a head-on collision of the electron with a
(realistic) finite nucleus, straight penetration and ensuing bilateral
oscillations result---``Coulomb oscillator.'' In calculations of the energy $%
E_{n0}$ of $l=0$ orbits, the minuscule influence from the nucleus' finite
size (part of the Lamb shift) is negligible for all practical purposes. In
this respect the conceptually different unilateral oscillation at one side
of a point nucleus (Coulomb repellor) and the bilateral Coulomb oscillator
through a finite nucleus are calculationally equivalent.

A windfall from the notion of an electron's (bilateral) Coulomb oscillation
through a finite nucleus is a natural interpretation of the hyperfine
splitting of $l=0$ energy levels---the so-called Fermi contact term--- 
\begin{equation}
\Delta E_{hfs}=-\frac{8\pi }{3}<\mathbf{\mu }_{N}\bullet \mathbf{\mu }%
_{el}>|\psi (0)|^{2}.  \tag{24}
\end{equation}

\noindent Here $\psi (0)$ is the spatial wave function of an $l=0$ electron
at the position of the nucleus and $<...>$ denotes the quantum-mechanical
coupling of nuclear and electron spin. Virtually all quantum mechanics texts
attribute the Fermi contact term to electron density $\psi \psi ^{*}$ near,
or \textit{``at''} the nucleus. A detailed inspection of scales shows,
however, that it arises not from outside ``contact'' of the electron with
the nucleus but only from its presence \textit{within} (Bucher 2000).
Independently derived, this interpretation is consistent with the concept of
the Coulomb oscillator.

\section{$\protect\sqrt{l(l+1)}$ CORRECTION}

Aside from the conceptual oversight of the Coulomb oscillator, a subtle
mathematical error occurred in Sommerfeld's transition of angular
quantization from 2D to 3D, Eqs. (8, 15, 16). The correct angular quantum
sum of the 3D quantization, $n_{\theta }+n_{\varphi }$, obtained from adding
the right-hand side of Eqs. (15, 16), agrees only accidentally with the
angular quantum number $n_{\alpha }$ of the 2D quantization in Eq. (8)
because of the \textit{double degeneracy} of the quantized Kepler problem,
that is, the dependence of the orbit energy on only the quantum sum $n$ but
not on the other quantum numbers, $l$ and $m_{l}$ (Schwarzschild 1916,
Epstein 1916). Sommerfeld's quantization by degrees of freedom, Eqs.
(14-16), was patterned after the treatment of conditionally periodic systems
with action-angle variables. It is known that spatial variables that lead to
double degeneracy---here, $r,$ $\theta $ and $\varphi $---are not quite
appropriate and can give rise to errors (Joos 1959). (Instead, sophisticated
orbit parameters, known in astronomy as Delaunay elements, have to be
employed.)

\includegraphics[width=5in]{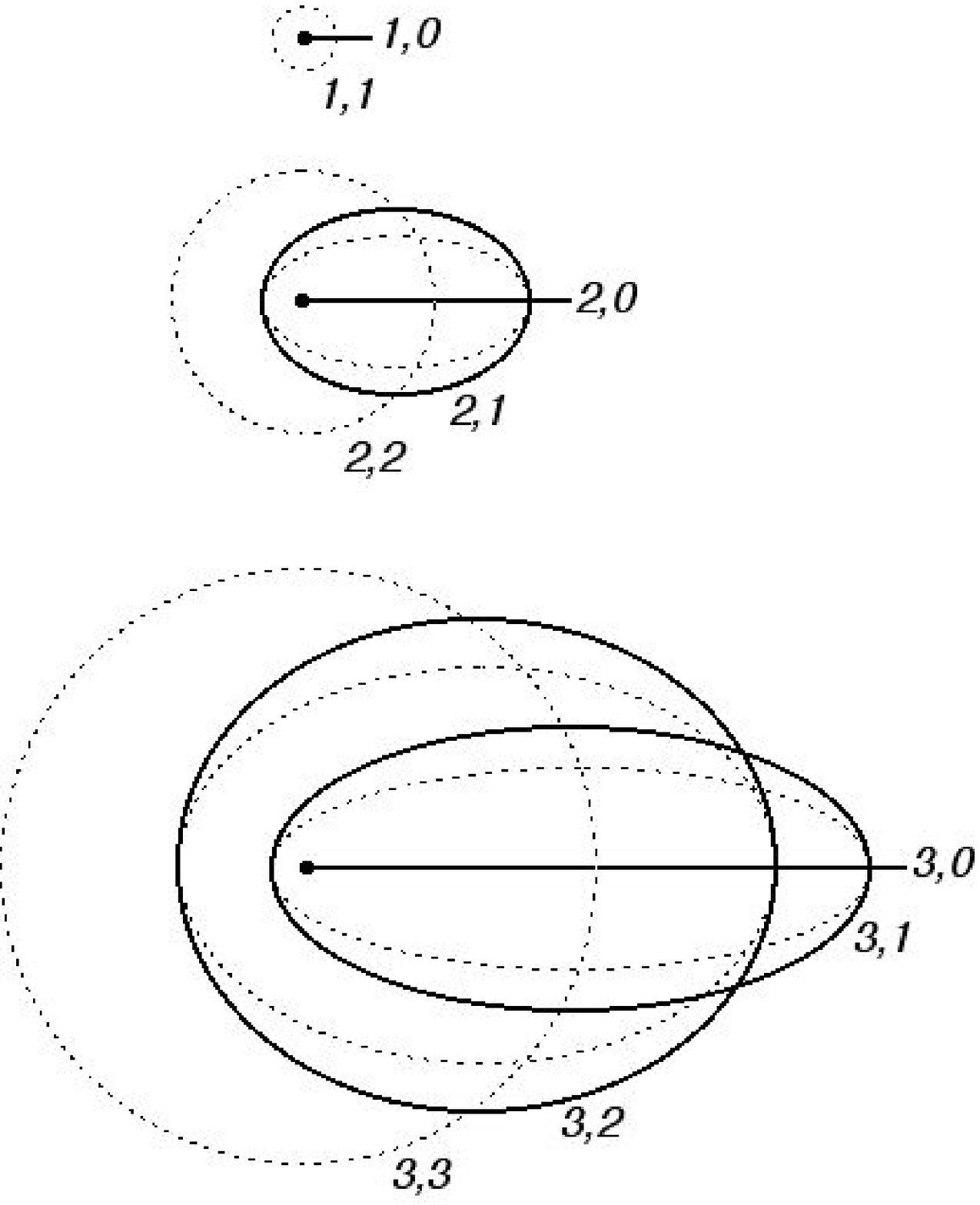}

\begin{quote}
Fig. 11.  Confocal display of historical (dashed) and correct $nl$ Sommerfeld orbits (solid lines) about a point nucleus \end{quote}\bigskip\

In the present case, the sum of the left-hand side of Eqs. (15, 16) must not
be identified with the action integral of the 2D orbit, Eq. (8). Experiment shows that the angular momentum---a constant of the motion---has the correct
magnitude $L_{nl}/h=\surd l(l+1)$, with angular quantum numbers ranging 
\begin{equation}
l=0,1,...,n-1,  \tag{25}
\end{equation}

\noindent instead of $L_{nl}/h$ $=l$ with $l=1,2,...,n,$ as Eq. (8)
erroneously implies. (This result agrees with quantum mechanics. However, no
derivation from quantization of the proper Delaunay elements has been
achieved yet.) Consistently, the correct length of semiminor axis must be 
\begin{equation}
b_{nl}=n\sqrt{l(l+1)}r_{B}  \tag{26}
\end{equation}
\noindent instead of Eq. (11). Figure 11 shows Sommerfeld ellipses with
correct widths, Eq. (22), including the (unilateral) Coulomb oscillator, in
comparison with the contradictory historical orbits. An important
consequence of this correction is the \textit{exclusion} of \textit{circular}
(Bohr) orbits, $l<n$. Another consequence is a corresponding modification of
the tilt angle, $\tau _{m_{l}}=arccos[m_{l}/\surd l(l+1)]$, instead of Eq.
(18). Therefore, a tilted Sommerfeld orbit can never lie in the equatorial
plane.

With the Coulomb oscillator included and the magnitude of the angular
momentum (and width of the tilted ellipses) corrected, the new Sommerfeld
orbits agree in energy, angular momentum and magnetic response with quantum
mechanics. Moreover, the time-average \textit{orbit size}, expressed through
elliptical semiaxes, Eqs. (10, 26), agrees with the corresponding results
from quantum-mechanical wave functions (Bucher et al. 1998), 
\begin{equation}
<r_{nl}>_{t}\ =\frac{3a_{nl}^{2}-b_{nl}^{2}}{2a_{nl}}=\frac{3n^{2}-l(l+1)}{2}
r_{B}=\int \psi ^{*}r\psi d^{3}r.  \tag{27}
\end{equation}

\noindent This orbit \textit{average} is readily visualized with the
corresponding $nl$ Sommerfeld ellipse. By Kepler theory, the average \textit{%
\ position} $P$ of an orbit lies halfway between the ellipse's center $C$
and empty focus $F^{\prime }$ (see Fig. 12). On the other hand, the distance
from the nuclear focus $F$ to point $D$, obtained by extending in Fig. 12
position $P$ parallel to the minor axis to the ellipse's perimeter, is the
orbit's average \textit{size}, $FD=$ $<r_{nl}>$, Eq. (27). A simple,
but  important case is the average size of the $H$ ground state, $nl=(1,0)$.
Equation (27) gives $<r_{10}>=\frac{3}{2}r_{B}$ instead of $1r_{B}$ as the
Bohr model implies, Eq. (2). This result is easily illustrated by the
distance $FD=FP$ on the (unilateral) rectilinear $(1,0)$ Coulomb-oscillator
orbit of major-axis length $2r_{B}$, Eq. (10).\bigskip\bigskip\bigskip\bigskip\bigskip\bigskip\bigskip

\includegraphics[width=3in]{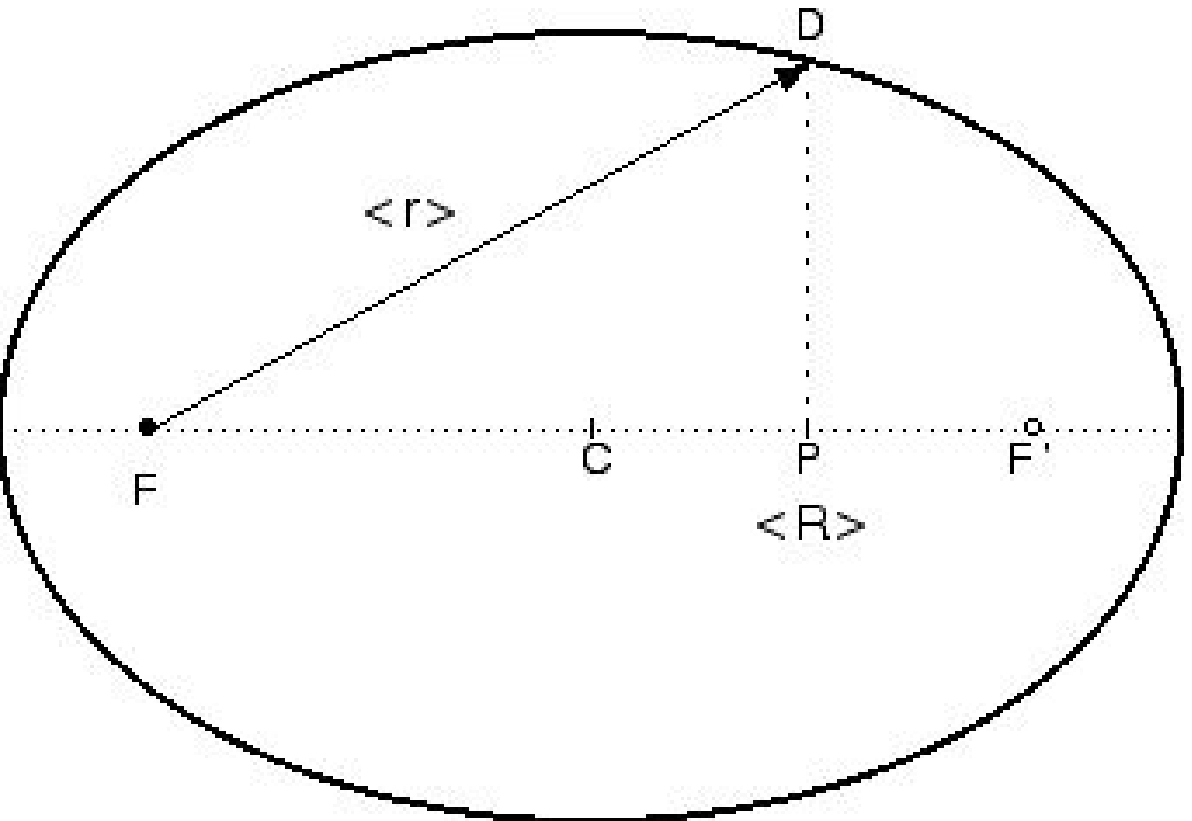}

\begin{quote}
Fig. 12.  Average distance $<r>=FD$ and average position $<\mathbf{R}>=P$ of an electron in a Sommerfeld orbit\end{quote}

The pathological aspect of the Sommerfeld quantization that affects the orbital
momentum (and elliptical minor axis), also affects the the fine-structure
formula, Eq. (19), where $l=1,2,...,n$. Because of the $\surd l(l+1)$
correction with $l=0,1,...,n-1$, the case $l=n$ (circular orbit) is no
longer admissible. On the other hand, the case $l=0$ (Coulomb oscillator)
renders the radical in Eq. (19) imaginary [or Eq. (20) divergent] and thus
physically meaningless. [The pathological behavior of Eqs. (19, 20) can, of
course, not be used to justify the exclusion of $l=0$ in the Sommerfeld
model since Sommerfeld\textit{\ explicitly excluded} this ``unphysical''
case in the derivation of his fine-structure formula.]

A modified fine-structure formula for the permitted range of angular quantum
numbers, $l=0,1,...,n-1$, is 
\begin{equation}
E_{nl}=m_{0}c^{2}\left\{ \left[ 1+\frac{\alpha ^{2}}{\left( l+\sqrt{%
(n-l)^{2}-\alpha ^{2}}\right) ^{2}}\right] ^{-\frac{1}{2}}-1\right\} , 
\tag{28}
\end{equation}

\noindent It differs from Sommerfeld's formula, Eq. (19), by an interchange
of the radial quantum number $n_{r}=n-l$ with the angular quantum number $l$
but yields the same set of energy values. [No derivation of Eq. (28) from
quantization of Delaunay elements has been achieved yet.] The remarkable
agreement of Sommerfeld's fine-structure formula, Eq. (19), with experiment
arises from a serendipitous cancellation of two errors---the quantization
pathology and the faulty range of $l$.

\section{CONCLUSION}

The triple success of the orbit-based, non-relativistic Bohr-Sommerfeld
quantum theory is its correct description of the $H$ atom (and one-electron
ions) concerning (1) the energy levels $E_{n}$, (2) the orbital angular
momenta $L_{nl}$---if corrected as $L_{nl}^{2}=l(l+1)\hbar ^{2}$and with the
Coulomb oscillator included---and (3) the orbits' space quantization---with $%
(L_{nl})_{z}=$ $m_{l}\hbar $. These three achievements are succinctly
represented by the corresponding (principal, angular and magnetic) quantum
numbers, $n,$ $l,$ $m_{l}$, and visualized by orbital ellipse
geometry---major axis $2a_{n}$, latus rectum $2\Lambda _{l}$, and tilt angle 
$\tau _{m_{l}}$ (Figs. 3, 4, 5).

The shortcomings of the old quantum theory lie in its neglect of three properties 
of particles---their spin, their wave nature and their quantum
statistics: Inclusion of spin---expressed by the spin quantum number $s$%
---is necessary to account for the \textit{total} (instead of only orbital)
angular momentum. The wave nature---expressed in Schr\"{o}dinger's
formulation of quantum mechanics by the wave function $\psi $---is necessary
to account for the particles' spatial (probability) distribution, $\psi \psi
^{*}=|\psi |^{2}$. Quantum statistics becomes relevant for systems with
identical (indistinguishable) particles---atoms, molecules, crystals.

The nagging problems and contradictions during the last years of the old
quantum theory were not caused by the orbit concept, as Pauli, Heisenberg
and Born suspected but, first and foremost, by the deliberate \textit{%
neglect of the Coulomb oscillator.} To be sure, ignorance of the electron
spin compounded the difficulties with the anomalous Zeeman effect, and
ignorance of quantum statistics hampered progress with the $He$ atom. All
those problems were readily solved after the discovery and elaboration of
matrix mechanics (Heisenberg 1925, Born et al. 1925), and more conveniently,
after the discovery of wave mechanics (Schr\"{o}dinger 1926)---consistent,
equivalent theories, collectively called quantum \textit{mechanics}.
Naturally, the old quantum theory was dropped. It is understandable that the
fathers of quantum mechanics kept pointing out the shortcomings of the old
quantum theory, rather than seek common ground, to highlight the
revolutionary character of their discoveries. (Not surprisingly, the
vindication of the old quantum theory's inherent Coulomb-oscillator concept
by quantum mechanics, $l=0$, went unnoticed: The radial component of a 
quantum-mechanical $s$ electron, with its finite---even maximum---probability 
density in the nucleus, can be regarded as a Coulomb oscillator, besides its 
wave-nature supplement by an exponential ``tunnel tail" beyond its classical region.) 
From a psychological perspective, Pauli's failure to solve the $H_{2}^{+}$ problem correctly in
his PhD thesis must have been a blow to that proud young man's self-esteem
and may well have caused his subsequent resentment toward the orbit concept.
[For the rest of his career, Pauli avoided, in word or print, reference to
his doctoral dissertation (Enz 2002).]

Sommerfeld, a parent of the atomic models, never recognized what went wrong
with his brain child. After the advent of quantum mechanics he \textit{%
redefined} the angular quantum number (Sommerfeld 1931) from previously $%
n_{\alpha }\equiv l$ and in the range $l=1,2,...,n$, Eqs. (17, 11), to 
\begin{equation}
n_{\alpha }\equiv l+1,\quad (l=0,1,...,n-1),  \tag{29}
\end{equation}

\noindent with $l$ now in the quantum-mechanical range. In this way he
retained both the old theory's exclusion of the Coulomb oscillator and
inclusion of the circular Bohr orbit---along with all its problems. Giving
in to quantum mechanics' paradigm of abstraction, he considered its \textit{%
formal} solution, $l=0$, a ``more satisfying justification'' for maintaining
his dictum $n_{\alpha }\neq 0$ than his former, orbit-based motivation of
avoiding electron collision with the nucleus. Does this reveal a general
psychological tendency? As a result of a \textit{visualizable} (here,
orbit-based) theory with celestial analogy, the solution $l=0$ is rejected
as an apparent contradiction (electron-nucleus collision). But as the result
of an \textit{abstract} (unvisualizable) theory, the same solution, $l=0$,
is acceptable.

Born, Heisenberg and Pauli all developed large schools where they trained
virtually all quantum physicists of their era. Naturally, their
collaborators, students and disciples adopted the masters' low opinion of
the old quantum theory and passed it on once they were teachers
themselves---a process multitudinously repeated to the present time. In this
way the old quantum theory acquired and maintained its low reputation of
being conceptually na\"{i}ve, logically inconsistent, and calculationally
impotent.

Is the old quantum theory really that bad? No. With the Coulomb oscillator
and the $\surd l(l+1)$ correction included, it is consistent with quantum
mechanics (albeit incomplete) and able to overcome its main historical
stumbling blocks (with solutions exact for $H$, approximate for $H_{2}^{+}$,
and plausible for $He$). This raises the question of whether quantum theory
would have developed differently if the Coulomb oscillator had been included
in the Sommerfeld model from the start. A different development is very
likely. Nevertheless, the old quantum theory would have fallen, probably
somewhat later, due to its oversimplified, unrealistic spatial description
of the atom, or more generally, its failure to account for the wave nature
of matter. Conceivably, it might have morphed, by combination with the
matter-wave hypothesis (de Broglie 1922), into some orbit-wave model, until
superceded---but also partly validated---by Schr\"{o}dinger's wave
mechanics. On the other hand, it is likely that the old quantum theory's
contradictions did accelerate the development of matrix mechanics by making
young Pauli and Heisenberg sufficently angry (Serwer 1977), and mature Born
desperate enough, to advance their radical, abstract theories. These are, of
course, conjectures and anybody's guess, affected by personal preference.

The remaining appeal of the Bohr-Sommerfeld model to \textit{some}
teachers---and virtually all students---is its visual, pedagogical asset
which complements quantum mechanics' abstraction. Combining both theories
brings out the best of two worlds: visualization and physical intuition
along with formal rigor and computational power. (The visual aspects of the
Bohr-Sommerfeld model---the connection of orbit parameters with quantum
numbers---can be traced in the wave functions of quantum mechanics. But that
will be another story.) Emphasizing the difference between quantum mechanics
and the old quantum theory may have been appropriate in the late 1920s but
it serves little purpose today. Seeking instead common ground benefits the
learning---and understanding---of quantum theory.

\section{APPENDIX: AUTOBIOGRAPHICAL NOTE}

How did I get involved with the old quantum theory, eight decades after its
fall? My involvement has deep roots and grew in a roundabout way. In my
doctoral thesis at the University of Frankfurt, Germany, I had studied
interatomic forces in silver halides---including $AgBr$, the traditional
photographic material---in comparison with the much better understood alkali
halides, like $NaCl$---table salt (Bucher 1984). The purpose was to employ
such forces in calculations of crystal defects in order to elucidate the
early stages of the photographic process (latent image). Information about
certain lattice defects, such as atoms in an ionic crystal or an electron
trapped at a vacancy ($F$ center), can be extracted from electron spin
resonance (ESR) experiments. The ESR signal (hyperfine interaction) is
proportional to the square of the probing electron's wave function, $|\psi
(r_{N})|^{2}$, at the position of a nucleus, $r_{N}$. When, as a
post-doctoral fellow at the University of Kansas, I calculated such wave
functions, Prof. Robert Friauf, my advisor, pointed out that ESR experiments
can be regarded as ``measuring'' the wave function at the nuclear position.
But when I searched the literature as to how close \textit{``at''} the
nucleus the probing electron was, I didn't find any clarification.

A few years later, while teaching hyperfine interaction in my
electrodynamics course at California State University, Fresno, I recognized
that the situation can be formulated as a local-field problem in terms of
atomic and nuclear scales, with averages over the latter. The formalism
revealed that for an $l=0$ electron, isotropic hyperfine interaction arises
only from the presence of the electron \textit{inside} its parent nucleus (Bucher 2000).

When around 2000 the opportunity arose to develop a conceptual
quantum-physics course at my university, I jumped at it. Studying, in
preparation, the old quantum theory's failure to give the correct multiplet
structure of the hydrogen atom, one sentence in Jammer's magisterial
treatment glared at me: ``The old quantum theory could never resolve this
inconsistency'' (Jammer 1966). I had a strong hunch how to resolve it---and
explain (isotropic) hyperfine interaction with the same stroke: by
straight-line oscillations of the electron through the nucleus. Its formal
development became a theory of the Coulomb oscillator (Bucher 2006). While
this extension of the Bohr-Sommerfeld theory was formally pleasing, the 
\textit{energy} of such Colomb oscillators was nothing new---it agreed with
those of the corresponding Bohr orbits (Bohr 1913a).

A different situation arose, however, for the hydrogen molecule ion, $%
H_{2}^{+}$. This entity falls into the gap between physics and chemistry.
Accordingly it is little known in either community. In essence, physics
stops at the hydrogen \textit{atom} and chemistry starts at the hydrogen 
\textit{molecule}, $H_{2}$. Hardly anybody seems responsible for $H_{2}^{+}$. Three notable exceptions are Pauling and Wilson's classical quantum text
(Pauling and Wilson 1935), Ruedenberg's review article (Ruedenberg 1962),
and Gutzwiller's chaos text (Gutzwiller 1990). It was not until I read Enz's
biography of Wolfgang Pauli that I became aware of the history of the
hydrogen molecule ion and Pauli's failure to solve it with the old quantum
theory (Enz 2002). Right away it occurred to me to apply the Coulomb
oscillator to $H_{2}^{+}$---an electron swinging in a straight line through
two separate protons. For large proton separations the results agreed
closely with the quantum-mechanical solution, as well as with another
semi-classical treatment (Strand and Reinhardt 1979). Less satisfactory
results for small proton separations alerted me to a second
solution---lateral electron oscillation in the bisectorial plane---that,
when taken into account, gave again good agreement with the exact,
quantum-mechanical solution (Bucher 2006). It took me some time, though, to
find justification for my double-solution procedure in the little-known
Einstein quantization condition (Einstein 1917).

One major problem was left where the old quantum theory had failed---the
helium atom. I made various, unsuccessful attempts of the Coulomb-oscillator
approach. But when I finally succeeded with contra-linear oscillations, I
was dumbfounded to obtain the same ground-state energy as Bohr with his
original, two-seat roundabout model of $He$ (Bohr 1913b). However,
subsequent explorations of both crossed and contra-linear oscillations
yielded on the one hand a ground-state energy that is comparable in accuracy
with the quantum-mechanical solution and, on the other hand, revealed Bohr's
historical value as the quantum-statistically forbidden triplet ground state
of $He$ (Bucher 2007).

\section{\noindent ACKNOWLEDGMENTS}

I thank Kurt Estel, Duane Siemens, Preston Jones, and most of all, Ernst
Mohler, for their patience to listen for years to my tale of the Coulomb
oscillator and for giving me valuable feedback. I also thank Prof. Fritz
Siemsen of the University of Frankfurt for his hospitality during my
sabbatical stay.

\section{REFERENCES}

\noindent Balmer JJ (1885) Notiz \"{u}ber die Spectrallinien des
Wasserstoffs. Verhandlungen der Naturforschenden Gesellschaft in Basel
7:548-560

\noindent Bohr N (1913a) On the constitution of atoms and molecules: Binding
of electrons by positive nuclei. Phil. Mag. 26:1-25

\noindent Bohr N (1913b) On the constitution of atoms and molecules: Systems
containing only a single nucleus. Phil. Mag. 26:476-502

\noindent Bohr N (1913c) On the constitution of atoms and molecules: Systems
containing several nuclei. Phil. Mag. 26:859-875

\noindent Bohr N (1923) Linienspektren und Atombau. Annalen der Physik
71:228-288

\noindent Boltzmann L (1868) Studien \"{u}ber das Gleichgewicht der
lebendigen Kraft zwischen bewegten materiellen Punkten. Wiener Berichte
58:517-560

\noindent Born M (1926) Zur Quantenmechanik der Sto\ss vorg\"{a}nge. Z.
Phys. 37:863-867

\noindent Born M, Heisenberg W (1923) Die Elektronenbahnen im angeregten
Heliumatom. Z. Phys. 16:229-243

\noindent Born M, Heisenberg W, Jordan P (1925) Zur Quantummechanik II. Z.
Phys. 35:557-615

\noindent Born M, Pauli W (1922) \"{U}ber die Quantelung gest\"{o}rter
mechanischer Systeme. Z. Phys. 10:137-158

\noindent Broglie de L (1923) Ondes et quanta. Comptes Rendus 177:507-510

\noindent Bucher M (1984) Interaction potentials for AgCl and AgBr. Phys.
Rev. B 30: 947-956

\noindent Bucher M, Elm D, Siemens DP (1998) Average position in Kepler
motion. Am. J. Phys. 66:929-930

\noindent Bucher M (2000) The electron inside the nucleus: An almost
classical derivation of isotropic hyperfine interaction. Eur. J. Phys.
21:19-22

\noindent Bucher M (2006) Coulomb oscillations as a remedy for the hydrogen
atom and molecule ion. arXiv:physics/0605258v2

\noindent Bucher M (2007) Coulomb oscillations as a remedy for the helium
atom. arXiv:physics/0705.4321v2

\noindent Charlier CL (1907) Die Mechanik des Himmels. Veit, Leipzig

\noindent Dalton J (1842) A new system of chemical philosophy. Bickersstaff,
London

\noindent Dirac PAM (1928) The quantum theory of the electron. Proc. Roy.
Soc. London (A) 117:610-624; 118:351-361

\noindent Ehrenfest P (1916) Adiabatische Invarianz und Quantentheorie.
Annalen der Physik 51:327-352

\noindent Einstein A (1905) \"{U}ber einen die Erzeugung und Verwandlung des
Lichtes betreffenden heuristischen Gesichtspunkt. Annalen der Physik
17:132-148

\noindent Einstein A (1906) Eine neue Bestimmung der Molek\"{u}ldimension.
Annalen der Physik 19:289-305

\noindent Einstein A (1907) Die Plancksche Theorie der Strahlung und die
Theorie der spezifischen W\"{a}rme. Annalen der Physik 22:180-190

\noindent Einstein A (1909) \"{U}ber die Entwicklung unserer Anschauungen
\"{u}ber das Wesen und die Konstitution der Strahlung. Phys. Z. 10:817-825

\noindent Einstein A (1917) Zum Quantenansatz von Sommerfeld und Epstein.
Verh. Dtsch. Phys. Ges. 19:82-92

\noindent Enz CP (2002) No time to be brief: A scientific biography of
Wolfgang Pauli. Oxford University Press, pp. 63-74.

\noindent Epstein PS (1916) Zur Quantentheorie. Annalen der Physik 51:168-188

\noindent Gordon W (1928) Die Energieviveaus des Wasserstoffatoms nach der
Diracschen Quantentheorie des electrons. Z. Phys. 48:11-14

\noindent Gutzwiller MC (1990) Chaos in classical and quantum mechanics.
Springer, New York, p. 33.

\noindent Heisenberg W (1925) \"{U}ber quantentheoretische Umdeutung
kinematischer und mechanischer Beziehungen. Z. Phys. 33:879-893

\noindent Heisenberg W (1927) \"{U}ber den anschaulichen Inhalt der
quantentheoretischen Kinematik und Mechanik. Z. Phys. 43:172-198

\noindent Ishiwara J (1915) Die universelle Bedeutung des 
Wirkungsquantums. Tokyo Sugaku Buturigakkawi Kizi 8:106-116

\noindent Jacobi CGJ (1866) Vorlesungen \"{u}ber Dynamik. Reimer, Berlin

\noindent Jammer M (1966) The conceptual development of quantum mechanics.
McGraw-Hill, New York, p. 193.

\noindent Jaworski BM, Detlaf AA (1972) Physik griffbereit. Vieweg,
Braunschweig, p. 718.

\noindent Joos G (1959) Lehrbuch der Theoretischen Physik. Akad.
Verlagsgesellsch., Frankfurt (11. Auflage), p. 127.

\noindent Kirchhoff GR (1859) \"{U}ber den Zusammenhang zwischen Emission
und Absorption von Licht und W\"{a}rme. Monatsberichte Akad. Wiss. Berlin,
Dez.

\noindent Kramers HA (1923) \"{U}ber das Modell des Heliumatoms. Z. Phys.
13:312-341

\noindent Land\'{e} A (1921) \"{U}ber den anomalen Zeemaneffekt. Z. Phys.
5:231-241

\noindent Mach E (1896) Die Principien der
W\"{a}rmelehre---historisch-kritisch entwickelt. Veit, Leipzig

\noindent Malow JV (1999) Another useful Keplerian average distance---the
harmonic mean. Am. J. Phys. 67:645-647

\noindent Mayer JR (1867) Die Mechanik der W\"{a}rme. Cotta, Stuttgart

\noindent Maxwell JC (1859) On the dynamical theory of gases. Brit. Assoc.
Repts. Aberdeen 29:9

\noindent Mendel\'{e}ev DI (1869) Sootnoshenie svoistv s atomnym vesom
elementov. Zhurnal Russkeo Fiziko-Khimichenskoe Obshchestvo 1:60-77

\noindent Meyer L (1864) Die Moderne Theorie der Chemie. Hirzel, Leipzig

\noindent Michelson AA (1891) On the application of interference-methods to
spectroscopic measurements. Phil. Mag. 31:338-346

\noindent Niessen KF (1923) Zur Quantentheorie des
Wasserstoffmolek\"{u}lions. Annalen der Physik 70:129-134

\noindent Ostwald W (1902) Vorlesungen \"{u}ber Naturphilosophie. Veit,
Leipzig

\noindent Paschen F (1916) Bohrs Helium Linien. Annalen der Physik 50:901-940

\noindent Pauli W (1922) \"{U}ber das Modell des Wasserstoffmolek\"{u}lions.
Annalen der Physik 68:177-240

\noindent Pauli W (1925a) \"{U}ber den Einflu\ss\ der
Geschwindigkeitsabh\"{a}ngigkeit der Elektronenmasse auf den Zeemaneffekt.
Z. Phys. 31:373-385

\noindent Pauli W (1925b) \"{U}ber den Zusammenhang des Abschlu\ss es der
Elektronengruppen im Atom mit der Komplexstruktur der Spektren. Z. Phys.
31:765-785

\noindent Pauli W (1926) \"{U}ber das Wasserstoffspektrum von Standpunkt der
neuen Quantenmechanik. Z. Phys. 36:336-363

\noindent Pauling L, Wilson EB (1935) Introduction to quantum mechanics.
Dover, New York, pp. 135-136.

\noindent Planck M (1900) \"{U}ber irreversible Strahlungsvorg\"{a}nge.
Annalen der Physik 1:69-122

\noindent Poincar\'{e} H (1905) Le\c{c}on de M\'{e}canique C\'{e}leste.
Gauthier-Villars, Paris

\noindent Ruedenberg K (1962) The physical nature of the chemical bond. Rev.
Mod. Phys. 34:326-276

\noindent Schr\"{o}dinger E (1926) Quantisierung als Eigenwertproblem.
Annalen der Physik 79:361-376

\noindent Schwarzschild K (1916) Zur Quantenhypothese. Berliner Berichte
1916:548-568

\noindent Serwer D (1977) Unmechanischer Zwang---Pauli, Heisenberg, and the
rejection of the mechanical atom. Historical Studies in the Physical
Sciences 8:189-256

\noindent Sommerfeld A (1911) Das Plancksche Wirkungsquantum und 
seine Bedeutung f\"{u}r die Molekularphysik. Phys. Z. 12:1057-1066

\noindent Sommerfeld A (1916) Zur Quantentheorie der Spektrallinien. Annalen
der Physik 51:1-94

\noindent Sommerfeld A (1931) Atombau und Spektrallinien. Vieweg,
Braunschweig, Vol. I, 5th revised ed., p. 126

\noindent Strand MP, Reinhardt WP (1979) Semiclassical quantization of the
low lying electronic states of $\mathit{H}_{\mathit{2}}^{\mathit{+}}$. J.
Chem. Phys. 70:3812-3827

\noindent Uhlenbeck GE, Goudsmit SA (1925) Ersetzung der Hypothese vom
unmechanischen Zwang durch eine forderung bez\"{u}glich des inneren
Verhaltens jedes einzelnen Electrons. Naturwissenschaften 13:953-954.

\noindent Van Vleck JH (1922) The normal helium atom and its relation to the
quantum theory. Phil. Mag. 44:812-869

\noindent Wilson W (1915) The quantum theory of radiation and line spectra. Phil. Mag. 29:795-802

\end{document}